\documentclass[%
aps,%
prb,
twocolumn,
superscriptaddress,%
preprintnumbers,%
byrevtex,%
floatfix%
]{revtex4}

\usepackage{dcolumn}
\usepackage{bm}
\usepackage{ifthen}
\usepackage[usenames]{color}
\usepackage{amssymb}
\usepackage{amsfonts}
\usepackage{amsmath}
\usepackage[dvips]{graphicx}

\newcommand{\beann} {\begin{eqnarray*}}
\newcommand{\eeann} {\end{eqnarray*}}
\newcommand{\bea} {\begin{eqnarray}}
\newcommand{\eea} {\end{eqnarray}}

\newcommand{\eg}{\textit{e.g.\,}}
\newcommand{\lrb} {\left(}
\newcommand{\rrb} {\right)}
\newcommand{\vect} {\mathbf}

\begin{document}
\date{\today}

\title{Vibrational modes and low-temperature thermal properties 
\\of graphene and carbon nanotubes: A minimal force-constant model}

\author{Janina Zimmermann}
\email{janina.zimmermann@iwm.fraunhofer.de}
\affiliation{Fraunhofer Institute for Mechanics of Materials IWM, D-79108 Freiburg, Germany}
\affiliation{Institute for Theoretical Physics, University of Regensburg, D-93040 Regensburg, Germany}
\author{Pasquale Pavone}
\affiliation{Institute for Theoretical Physics, University of Regensburg, D-93040 Regensburg, Germany}
\author{Gianaurelio Cuniberti} 
\affiliation{Institute for Theoretical Physics, University of Regensburg, D-93040 Regensburg, Germany}
\affiliation{Institute for Materials Science and Max Bergmann Center of Biomaterials, Dresden University of Technology, D-01062 Dresden, Germany}

\begin{abstract}
We present a phenomenological force-constant model developed for the description of lattice dynamics of $sp^2$ hybridized carbon networks.
Within this model approach, we introduce a new set of parameters to calculate the phonon dispersion of graphene by fitting the \emph{ab initio} dispersion.  
Vibrational modes of carbon nanotubes are obtained by folding the 2D dispersion of graphene and applying special corrections for the low-frequency modes.
Particular attention is paid to the exact dispersion law of the acoustic modes, which determine the low-frequency thermal properties and reveal quantum size effects in carbon nanotubes.
On the basis of the resulting phonon spectra, we calculate the specific heat and the thermal conductance for several achiral nanotubes of different diameter.
Through the temperature dependence of the specific heat we demonstrate that phonon spectra of carbon nanotubes show one-dimensional behavior and that the phonon subbands are quantized at low temperatures.
Consequently, we prove the quantization of the phonon thermal conductance by means of an analysis based on the Landauer theory of heat transport. 
\end{abstract}


\maketitle
\section{Introduction}

Phonons play a fundamental role in the physics and the characterization of graphene and carbon nanotubes.
Phenomena such as charge,~\cite{Nemec2006,Nemec2008a,Nemec2008b} spin~\cite{Krompiewski2004} and heat transport,~\cite{Balandin2008,Gheorghe2005,Berber2000,KSaito2007} infrared and Raman spectra,~\cite{Kim2005,Ferrari2006,Rao1997} electron-phonon scattering~\cite{Woods2000,Dubay2002,Piscanec2004,Heid2004,Lazzeri2005,Gutierrez2006,Bonini2007} and its related effects as superconductivity~\cite{Tang2001} and resistivity~\cite{Suzuura2002,Mariani2008} can be understood, in most situations, only with a detailed knowledge of the phonon spectrum.
In particular, much effort has been done for determining thermal and transport properties, which closely depend on the vibrational modes.
The most striking results of experimental and theoretical research in this domain are the observation of the quantization of the phonon band structure through an analysis of the specific heat,~\cite{Hone2000} the discovery of ballistic phonon transport~\cite{Rego1998} and the measurement of the quantum of thermal conductance in a nanowire.~\cite{Schwab2000} 
In technological applications such as nanotube-based electronic devices, thermal properties are of central importance for understanding and controlling heat dissipation and self-heating effects.~\cite{Ouyang2006} Efficient thermal management is required for ensuring the performance and stability of the devices.

Much experimental work has been done for measuring vibrational spectra~\cite{Siebentritt1997,Maultzsch2004,Ferrari2006,Rao1997} and for detecting and controlling the phonon population of isolated nanotubes.~\cite{LeRoy2004} 
The best known feature of experimental data is the strong Raman-active radial breathing mode (RBM), which is often used for the characterization or identification of different nanotubes in a sample.~\cite{Jorio2001,Telg2004} 
From a theoretical point of view, phonon modes of graphene have been studied either by effective models~\cite{Benedek1993,Jishi1993} or by \emph{ab initio} calculations.~\cite{Pavone1996,Rubio1999,Dubay2003,Maultzsch2004,Wirtz2004,Piscanec2004,Mounet2005}
Several models have been proposed for the lattice dynamics of carbon nanotubes, ranging from zone folding and force-constant models,~\cite{Jishi1993,Saito1998,Saito1998paper,Mahan2004,Kandemir2008} valence force-field models,~\cite{Popov1999,Popov2000,Gunlycke2008} and tight binding~\cite{Popov2006} to \emph{ab initio} calculations.~\cite{Rubio1999,Dubay2003,Heid2004,Liu2004}

In this work we concentrate on the well-established fourth-nearest-neighbor force-constant model by Jishi~\emph{et al.}.~\cite{Jishi1993} It was developed and optimized for graphene and subsequently also adapted to carbon nanotubes by Saito~\emph{et al.}.~\cite{Saito1998,Saito1998paper}
It has been reparametrized several times for graphene~\cite{Grueneis2002,Samsonidze2003,Wirtz2004} but, to our knowledge, no further calculations for carbon nanotubes have been done. While the original parameters were empirically determined by fitting experimental data of graphite, we propose a new parametrization to fit the \emph{ab initio} phonon dispersion of graphene.~\cite{HeidPrivat} With some corrections, we use this parametrization also for the calculation of the phonon dispersion of achiral carbon nanotubes.
Main attention is paid to the long-wavelength acoustic modes and to the controversial question of the dispersion law of the transverse acoustic (TA) or flexure mode.~\cite{Mahan2004}  
While Saito \emph{et al.}~\cite{Saito1998paper} obtain four linear-dispersing acoustic modes ($\omega\propto q$), we obtain two linear modes and a doubly degenerate quadratic dispersing flexure mode ($\omega\propto q^2$). 
The quadratic dependence of the flexure modes of carbon nanotubes, predicted by 
continuum models~\cite{Suzuura2002,Goupalov2005,Chico2006} and obtained by several \emph{ab initio} calculations,~\cite{Dubay2003,Liu2004,Rubio1999} 
has been reproduced only by few force-constant models.~\cite{Gartstein2004,Popov2000,Mahan2004}
One of the latter (the work of Mahan and Jeon~\cite{Mahan2004}) pursued a detailed study of the symmetry rules that lead to a quadratic flexure mode and achieved the correct behavior with a three-parameter spring and mass model. 
It is chosen for comparisons throughout this paper.

We concentrate on an accurate description of the acoustic phonons since they
allow to predict and interpret several low-temperature thermodynamic properties and to prove quantum size effects in carbon nanotubes.
Using our force-constant model we calculate the specific heat and the thermal conductance of carbon nanotubes of different diameter and chirality. Although the exact dispersion law of the acoustic modes is apparently irrelevant for the quantized thermal conductance,~\cite{Mingo2005} the quadratic dispersion of the flexure modes results in a very different behavior of the low-temperature specific heat.~\cite{Popov2002}
Several experimental and theoretical studies have been achieved for determining the specific heat of carbon nanotubes and nanotube ropes. 
Most of them~\cite{Mizel1999,Hone2000,Dobardzic2003,Li2005} 
cover a wide temperature interval and provide approximate estimations of the power law of the $T$-dependence, but only few works~\cite{Popov2002,Lasjaunias2002} extend the temperature range down to $\sim$0.1~K and provide a more precise analysis. 
Popov~\cite{Popov2002} showed within a force-constant model that a $T^{0.5}$ dependence of the specific heat at very low temperatures can be directly related to the flexure modes.
Lasjaunias~\emph{et al.}~\cite{Lasjaunias2002} determined experimentally on a sample of nanotube ropes that under $\sim$1~K dominates a $T^{0.62}$ dependence.
We study the exact power law including temperatures in the mK region and illustrate how it is correlated with the acoustic modes and the dimensionality of the system. As in Ref.~[\onlinecite{Hone2000}] we prove the 1D quantization of the phonon subbands in nanotubes.

Quantum size effects in carbon nanotubes are observed even in the thermal conductance. 
The thermal conductance of phonon wave\-guides in the ballistic, one-dimensional limit has been calculated by Rego~\emph{et al.}~\cite{Rego1998} using the Landauer formula and has been proved experimentally by Schwab~\emph{et al.}.~\cite{Schwab2000} 
Within the same formalism, we show that the phonon thermal conductance of carbon nanotubes is quantized and determine the thermal conductance quantum. Our results are in good agreement with theoretical calculations.~\cite{Watanabe2004,Mingo2005}

This paper is structured as follows. In Sec.\ II we provide a brief description of the system.
Section III deals entirely with vibrational properties: After a summary of lattice dynamics and the model approach in Sec.\ III A-B, we present our results for the phonon dispersions of graphene and carbon nanotubes in Sec.\ III C-D. In Sec.\ IV we consider thermal properties: The basic concepts are recalled in Sec.\ IV A, while Sec.\ IV B-C show our calculations of the low-temperature specific heat and the thermal conductance, respectively. Section V contains final remarks.

\section{System}

The typical honeycomb structure of graphene is defined by a 2D hexagonal lattice with 
a basis of two atoms, which we call atom $A$ and $B$. The lattice vectors are given by $\mathbf{a}_1=\left( \sqrt{3}a/2,a/2\right )$
and $\mathbf{a}_2=\left( \sqrt{3}a/2,-a/2\right )$ with the lattice constant $a=2.46$~\AA, as 
\begin{figure}[b]
\centerline{\includegraphics[width=6.8cm]{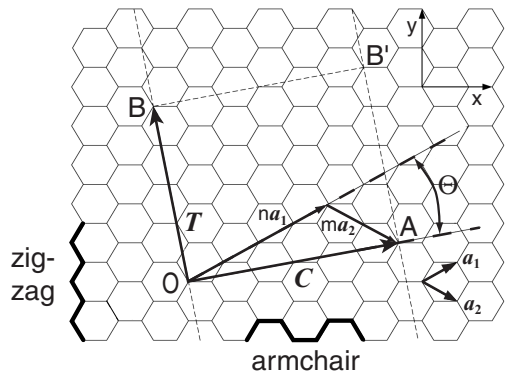}}
\caption{\label{fig:network}
The graphene honeycomb lattice with lattice vectors $\mathbf{a}_1$ and $\mathbf{a}_2$. 
A carbon nanotube can be constructed by rolling up
the graphene sheet along $\mathbf{C}$ so that points O and A coincide (as well as B and B' do). 
$\Theta$ denotes the chiral angle. (Figure taken from Ref.~[\onlinecite{Thune05}])
}
\end{figure}
illustrated in Fig.~\ref{fig:network}.

A carbon nanotube can be thought of as a single graphene sheet that is wrapped
into a seamless cylinder. It is common to define a circumferential vector and a vector parallel 
to the tube axis.~\cite{ReichTM04,Saito1998}
The first one, called chiral vector, is defined in terms of the unit vectors of graphene,
$\mathbf{C}=n\mathbf{a}_1+m\mathbf{a}_2$, and the sheet is rolled up in such a way that it becomes the 
circumference of the tube. The pair of integers $(n,m)$ uniquely defines a particular nanotube and thus provides 
a classification among nanotubes.
The translational vector $\mathbf{T}$ is perpendicular to $\mathbf{C}$ and reproduces the periodicity of the 
nanotube structure along the axis direction. It reads $\vect{T} = \lrb (2m +
n)/d_R\rrb \vect{a}_1 - \lrb (2n + m)/d_R\rrb \vect{a}_2$, being
$d_R$ the greatest common divisor of $(2m + n)$ and $(2n + m)$, as
$\vect{T}$ should be the smallest lattice vector in its direction.
Alternatively a nanotube can be defined also by its radius $R$ and the chiral angle $\Theta$, which is given 
by the chiral vector measured relative to the direction defined by $\mathbf{a}_1$. 
In this work we concentrate on the particular cases of zigzag ($\Theta=0$) 
and armchair ($\Theta=\pi/6$) nanotubes, which represent the class of achiral nanotubes.


\section{Vibrational properties}
\subsection{Lattice dynamics}\label{Sec:latticedyn}
To derive the equations of motion for the atoms we use Hamiltonian mechanics,
treating the atoms as point masses moving according to the laws of classical mechanics.
To describe the ion configuration, characterized by the instantaneous location of the atoms, 
we use the following notation for a crystal with a monoatomic basis
\bea \tilde{\mathbf{R}}_n(t)=\mathbf{R}_n+\mathbf{u}_n(t).
\eea
Hence at time $t$ the ion is located at $\tilde{\mathbf{R}}_n(t)$, while $\mathbf{R}_n$ is its 
equilibrium position.
In the limit of small displacements $\mathbf{u}_n$ of the atoms from their equilibrium position, 
the so-called \emph{harmonic approximation}, the equations of motion are a set of
coupled second order differential equations given by
\begin{equation}
M  \ddot{\mathbf{u}}_n = - \sum_{m} \Phi(\mathbf{R}_{n},\mathbf{R}_{m})\cdot \mathbf{u}_{m}
\label{Eq:motion}
\end{equation}
where $M$ is the mass of the constituent atom and $\Phi(\mathbf{R}_{n},\mathbf{R}_{m})$ is the 
$3\times 3$ \emph{force-constant tensor} that couples atom $n$ and $m$.
Due to lattice periodicity it is possible to search for solutions of Bloch-wave type
\bea \mathbf u_n^{\mathbf q}(t)=\mathbf A \,\textrm{e}^{\textrm{i}\mathbf{q}\cdot \mathbf{R}_n}\textrm{e}^{-\textrm{i}\omega t}
\label{Bloch:wave}
\eea
where $\mathbf A$ gives the amplitude of the mode, $\omega$ the frequency, and $\mathbf q$ the wave vector. Inserting Eq.~(\ref{Bloch:wave}) in~(\ref{Eq:motion}),
the equations of motion become
\bea \omega^2 M \mathbf{A}=\sum_m  \Phi(\mathbf{R}_{n},\mathbf{R}_{m})\,\textrm{e}^{\textrm{i}\mathbf{q}\cdot(\mathbf{R}_{m}-\mathbf{R}_{n})} \mathbf{A}.  
\eea
These can be written in the compact form
\bea \textrm{D}(\mathbf{q})\cdot\mathbf{A}=\omega^2\mathbf{A} 
\eea
where we introduced the discrete Fourier transform
\bea \textrm{D}(\mathbf{q})=\frac{1}{M} \sum_m \Phi(0,\mathbf{R}_{m})\,\textrm{e}^{\textrm{i}\mathbf{q}\cdot\mathbf{R}_{m}}.
\eea
The matrix $\textrm{D}(\mathbf{q})$ is called \emph{dynamical matrix} and is a hermitian and positive definite matrix. 
In order to obtain the eigenvalues $\omega^2(\mathbf q)$ and the eigenvectors $\mathbf A(\mathbf q)$ we have to solve the secular problem $\textrm{det\,D}(\mathbf q)=0$ for each $\mathbf q$ vector chosen according to the Born-von-K\'arm\'an periodic boundary conditions.
The generalization of the above described theory to non-monoatomic basis systems is easy and can be found, \eg, in Ref.~[\onlinecite{Roessler2004}].
Considering a three-dimensional system with $r$ atoms per unit cell, the dynamical matrix has $3r\times 3r$ components so that for each wave vector $\mathbf q$ there are $3r$ frequencies $\omega_s(\mathbf q)$, with $s=1,\ldots,3r$.

\subsection{Force-constant model}
A practical method of investigating vibrational properties of graphene and carbon nanotubes (CNTs) is given by phenomenological lattice-dynamical models. 
These try to construct the force-constant tensor starting by an analytic expression for the interaction energy of two or more carbon atoms,~\cite{Mahan2005,Mahan2004,Gartstein2004} or alternatively by approximating directly the interatomic force constants 
by fitting experimental data.~\cite{Jishi1982,Jishi1993}
Such empirical models are based on a few adjustable parameters and are able to provide reliable information that is complementary to that obtainable from more advanced methods. Indeed, an alternative tool is given by first principles or \emph{ab initio} calculations based on the quantum mechanical description of electrons.~\cite{Baroni2001,Pavone2001} This method does not rely on input from experimental informations and includes all relevant effects, providing accurate, experimentally-confirmed and therefore very predictive results, as has been shown, \eg, in Ref.~[\onlinecite{Giannozzi1991,Pavone1993}]. However, the computational effort is large, leading to several restrictions in particular for complex systems of considerable size.
The advantage of the phenomenological models consists in their simplicity and the possibility of fast application to almost every system.
In view of the aim of our work, force-constant models turn out as the best choice for two reasons: (i) they provide quick and reliable implementation for several CNTs of different diameter and chirality, (ii) they reproduce with a high level of accuracy especially the low-energy acoustic modes, which in turn determine almost entirely the low-temperature thermal properties of CNTs.

\begin{figure}[t]
\centerline{\includegraphics[width=5.5cm]{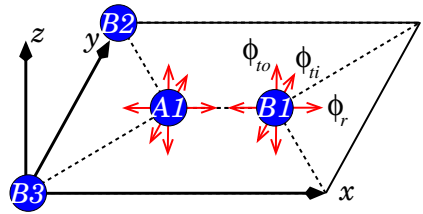}}
\caption{
An atom $A$ and its first nearest-neighbor atoms $Bp$ ($p=1,2,3$). $\phi_r,\, \phi_{ti},$ and $\phi_{to}$ represent forces in radial, in-plane, and out-of-plane direction.
\label{NN:forces}
}
\end{figure}

We calculate the phonon modes of graphene and carbon nano\-tubes using the force-constant model proposed by Saito \emph{et al.}.~\cite{Saito1998} This model consists in the direct parametrization of the diagonal real-space force constants including up to fourth nearest-neighbor interactions (4NNFC approach). This leads to a set of twelve adjustable parameters. The truncation after the fourth nearest neighbors is justified by the rapid decay of the force constants.~\cite{Dubay2003,Mounet2005}
In the 4NNFC approach the force-constant tensor describing the interaction between an atom and its $n$th nearest neighbor on an arbitrarily chosen axis (\eg, the \emph{x} axis) has diagonal form
\begin{equation}
\Phi=
\left ( \begin{array}{ccc}
    \phi_r^{(n)} & 0 & 0 \\
    0 & \phi_{ti}^{(n)} & 0 \\
    0 & 0 & \phi_{to}^{(n)}
    \end{array} \right )
\label{FCtensor:Dress}
\end{equation}
where $\phi_r^{(n)},\, \phi_{ti}^{(n)},$ and $\phi_{to}^{(n)}$ represent the force-constant parameters in the radial (bond-stretching), in-plane, and out-of-plane tangential (bond-bending) directions of the $n$th nearest neighbors.
The radial direction corresponds to the direction of the bonds and the two tangential directions are perpendicular to it, as illustrated in Fig.~\ref{NN:forces}.
The force-constant tensors for nearest-neighbor atoms of the same neighbor shell, that are not located on the \emph{x} axis, can be obtained by unitary rotation of the tensor of Eq.~(\ref{FCtensor:Dress}). 
The formalism is described accurately in Ref.~[\onlinecite{Saito1998}].

For example for first nearest neighbors ($n=1$) we obtain the force-constant tensor $\Phi^{(A,Bp)}$ between atom $A$ and its neighbor $Bp$ ($p=2,3$) by
\begin{equation}
\Phi^{(A,Bp)}=U_z^{-1}(\theta_p)\,\Phi^{(A,B1)}\,U_z(\theta_p)
\end{equation}
where $U_z(\theta_p)$ is a unitary rotation matrix around the \emph{z} axis
\begin{equation}
U_z(\theta_p)=
\left ( \begin{array}{ccc}
    \cos(\theta_p) & \sin(\theta_p) & 0 \\
    -\sin(\theta_p) & \cos(\theta_p) & 0 \\
    0 & 0 & 1
    \end{array} \right )
\label{Rotation:matrix}
\end{equation}
and $\theta_p$ is the angle defined by atom $B1$, $A$, and $Bp$.

The above mentioned tensors describe the interaction between atoms in the plane and for carbon nanotubes these have to be adapted because of the curvature of the walls. It is possible to generate the force-constant tensors for the atoms of the nanotube unit cell by rotation of the che\-mi\-cal bond from the two-dimensional plane of graphene to the three-dimensional coordinates of the nanotube.~\cite{Saito1998} It is possible to generate the force-constant tensors for all the atoms of the unit cell of a nanotube from those related to one single atom of the cell, \eg atom $A1$ of Fig.~\ref{NN:forces}. For atoms of type $A$ the tensors of atom $A1$ must be rotated by an angle $\Psi_i$ around the axis of the nanotube (here the \emph{y} axis)
%
\begin{equation}
\Phi^{(Ai,Bp)}=U^{-1}_y(\Psi_i)\,\Phi^{(A1,Bp-i+1)}\,U_y(\Psi_i).
\end{equation}
%
$\Psi_i$ is the polar angle between $A1$ and $Ai$ around the circumference.
If $(p-i+1)$ is negative or zero we use $(\frac{r}{2}+p-i+1)$ instead of it.
For atoms of type $B$ we must rotate the tensors of atom $A1$
first by $\pi$ around the \emph{z} axis, and then by $\Psi_i$ around the \emph{y} axis, as before
%
\begin{equation}
\Phi^{(Bi,Ap)}=U^{-1}_y(\Psi_i)\,U^{-1}_z(\pi)\,\Phi^{(A1,Bp-i+1)}\,U_z(\pi)\,U_y(\Psi_i)
\end{equation}
%
where $U_y(\psi)$ and $U_z(\psi)$ are unitary rotation matrices around the \emph{y}- and \emph{z} axis, analogously to that of Eq.~(\ref{Rotation:matrix}). 
The dynamical matrix is obtained by multiplying the force constant tensors obtained above by exp$(iq_znT)$, where $n$ is the number of the unit cell in which atom $A1$ is situated, and $T=|\mathbf T|$ is the modulus of the translational vector.

\subsection{Results for graphene}
In the calculation of the phonon-dispersion relation of graphene done by Saito \emph{et al.},~\cite{Saito1998} the force-constant parameters were empirically determined by fitting experimental data of graphite obtained by inelastic neutron scattering. We perform, instead, a parameter fit to the \emph{ab initio} dispersion 
relation of graphene calculated within density-functional perturbation theory by Bohnen and Heid.~\cite{HeidPrivat} 
The corresponding sets of force constants are listed in Table~\ref{tab:dr1}.

\begin{figure}[t]
\centerline{\includegraphics[width=6.7cm,height=10cm]{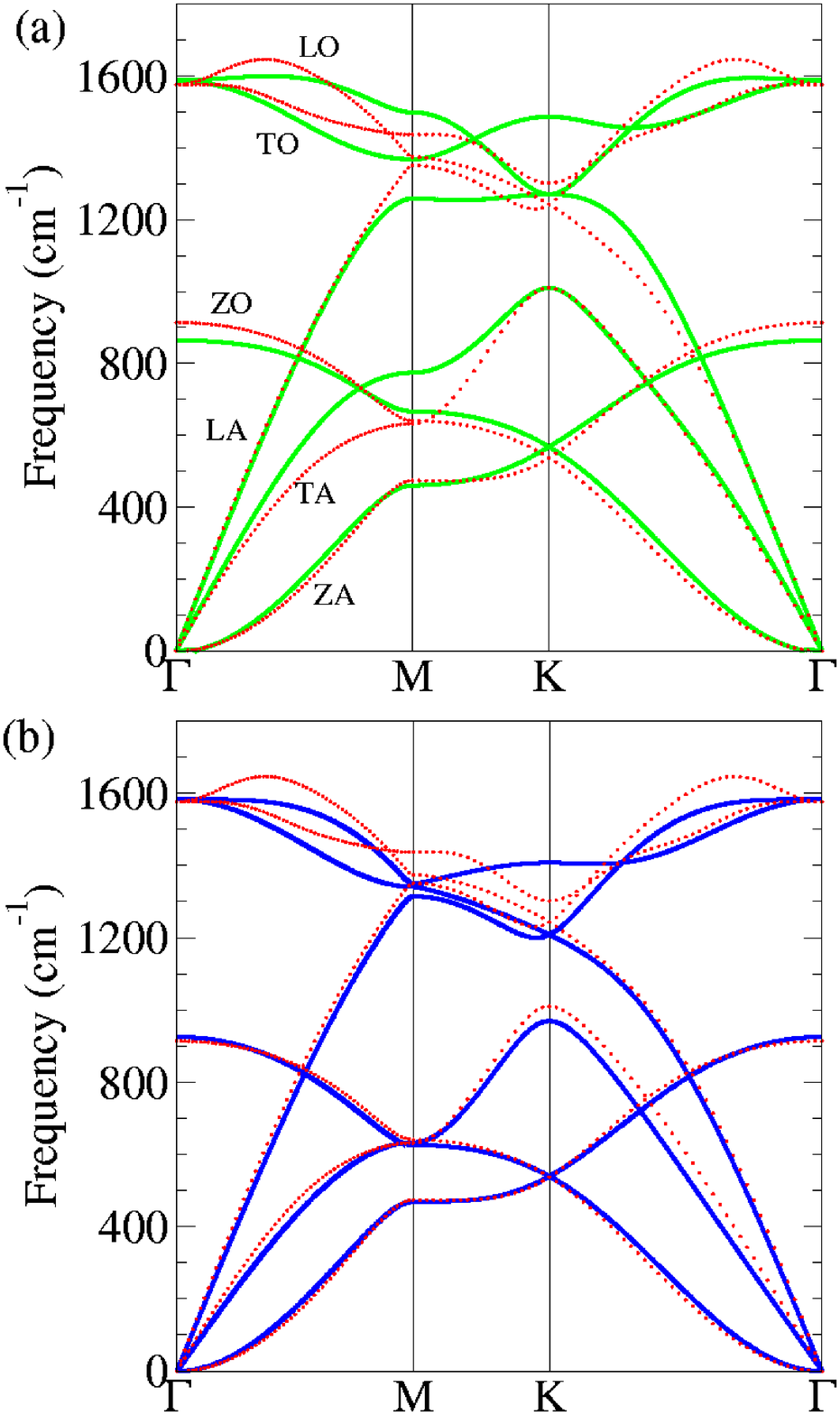}}
\caption{ 
Dispersion relation of graphene calculated with the fourth-nearest-neighbor model (4NNFC) (solid lines) in direct comparison with \emph{ab initio} calculations (dotted lines) of Bohnen and Heid.~\cite{HeidPrivat} (a) 4NNFC approach with the original parametrization of Saito \emph{et al.}; (b) 4NNFC approach with our parametrization. The corresponding sets of parameters are listed in Table~\ref{tab:dr1}.
\label{Disp:graphene}
}
\end{figure}

Precisely, the fitting process was performed through a matching of the force-constant tensor to those obtained by \emph{ab initio} calculations.~\cite{HeidPrivat} This procedure is however limited by the constraint of including up to fourth nearest neighbors in the force-constant tensor, as required by the model approach. 
We varied and optimized the force constants in order to fit as closely as possible the \emph{ab initio} phonon dispersion.
\begin{table*}[t]
 \begin{tabular}{p{0.2\textwidth}|p{0.1\textwidth}p{0.1\textwidth}p{0.1\textwidth}|p{0.1\textwidth}p{0.1\textwidth}p{0.1\textwidth}}
 \toprule
  & \multicolumn{3}{c|}{Parameters by Saito \emph{et al.}~(Ref.~[\onlinecite{Jishi1993}])} &\multicolumn{3}{c}{Our parametrization}    \\
 \multicolumn{1}{c|}{Neighbor shell} & $\phi_r^{(n)}$ &  $\phi_{ti}^{(n)}$ &  $\phi_{to}^{(n)}$ & $\phi_r^{(n)}$ & $\phi_{ti}^{(n)}$ & $\phi_{to}^{(n)}$ \\
\hline
 $1$st &   36.50   &  24.50 &  9.82  & 41.8  & 15.2  & 10.2  \\
 $2$nd &    8.80   &  -3.23 & -0.40  & 7.6   & -4.35 & -1.08 \\
 $3$rd &    3.00   &  -5.25 &  0.15  & -0.15 & 3.39  & 1.0   \\
 $4$th &   -1.92   &   2.29 & -0.58  & -0.69 & -0.19 & -0.55 \\
 \hline \hline
 \end{tabular}
 \caption{Force-constant parameters for graphene in units of $10^4~\textrm{dyn/cm}=10~\textrm{N/m}$.\label{tab:dr1}}
 \end{table*}
%
We follow Gartstein~\cite{Gartstein2004} choosing the in- and out-of-plane tangential force constants $\phi_{t}^{(n)}$ so as to satisfy
$\phi_{t}^{(1)}+6\phi_{t}^{(2)}+4\phi_{t}^{(3)}+14\phi_{t}^{(4)}=0$. This equality is required by the
rotational invariance of the graphene plane, and the original parameters of Saito and coworkers do not obey this rule.
Figure~\ref{Disp:graphene} shows the phonon-dispersion relation resulting from: (a) the original parametrization and (b) our new parametrization. Both curves are superposed to the \emph{ab initio} dispersion of Bohnen and Heid (dotted lines), for direct comparison. 
%
%

The dispersion relation of graphene comprises three acoustic (A) and three optical (O)
modes, which are either out-of-plane (Z), in-plane longitudinal (L) or transverse (T).
The acoustic ZA mode shows a $q^2$ energy dispersion near $\Gamma$ rather than the linear dispersion of the TA and LA mode, which is typical for acoustic modes in 3D solids. The quadratic dispersion is a characteristic feature of the phonon dispersion of layered crystals~\cite{Zabel2001} and can be explained as a
consequence of the $D_{6h}$ point-group symmetry of graphene.~\cite{Saito1998} Another
consequence of the symmetry are the linear crossings of the ZA/ZO
modes and the LA/LO modes at the K point. With respect to the phonon dispersion obtained by Saito \emph{et al.} and in comparison with first-principles results, our new parametrization yields a considerable improvement in the overall phonon dispersion. In particular the acoustic modes provide a remarkable good fit to \emph{ab initio} data.
The frequency values at high symmetry points $\Gamma$, M and K (listed in Table~\ref{tab:values}) differ only by up to 4\% from \emph{ab initio} data, with exception of the TO mode (6.7\% at M and 8\% at K), which will be discussed later.
 \begin{table}[t]
 \begin{tabular}{llccc}
 \hline
 \hline
  & Mode & Ref.~[\onlinecite{HeidPrivat}] & This work & Ref.~[\onlinecite{Mounet2005}]\\
  &      & \hspace*{0.1cm} \emph{Ab initio} LDA  \hspace*{0.1cm}       &           & \hspace*{0.1cm}  \emph{Ab initio} GGA \hspace*{-0.1cm} \\
 \hline
 $\Gamma$ &    ZO      &  914  &  925   &  881    \\ 
          &    LO/TO   &  1576 &  1583  &  1554   \\ 
  M\hspace*{0.3cm}     &    ZA      &  474  &  469   &  471    \\ 
          &    TA      &  632  &  626   &  626    \\
          &    ZO      &  641  &  633   &  635    \\
          &    LA      &  1351 &  1315  &  1328   \\
          &    LO      &  1375 &  1351  &  1340   \\
          &    TO      &  1437 &  1341  &  1390   \\ 
  K       &    ZA/ZO   &  538  &  539   &  535    \\
          &    TA      &  1010 &  969   &  997    \\ 
          &    LA/LO   &  1243 &  1208  &  1213   \\
	  &    TO      &  1302 &  1408  &  1288   \\	   	  	  	  		
 \hline 
 \hline
 \end{tabular}
 \caption{Phonon frequencies (cm$^{-1}$) of graphene at high symmetry points. \label{tab:values}}
 \end{table}
%
In the high-energy range, our parametrization leads to a qualitatively correct rearrangement of the LA and LO modes along the line M-K and an improvement concerning the crossing of the LO and TO branches along the $\Gamma$-M and $\Gamma$-K directions.
Nevertheless, there are still major divergencies from the \emph{ab initio} dispersion for the LO and TO mode. Neither parametrization reproduces the initial upward curvature, called overbending, of the LO branch away from $\Gamma$ that is observed in both the experimental data~\cite{Oshima1988,Aizawa1990,Siebentritt1997,Maultzsch2004} and in all published first-principles calculations.~\cite{Rubio1999,Pavone1996,Dubay2003,Maultzsch2004,Wirtz2004,Piscanec2004,Mounet2005}
Furthermore, the TO phonon at the K-point is significantly higher than in the \emph{ab initio} dispersion.
It is known from literature that the highest optical phonon branch is shaped by the effect of electron-phonon interactions, which results in a discontinuity in the frequency derivative at $\Gamma$ and K.~\cite{Piscanec2004} These discontinuities are called Kohn anomalies and are revealed by two sharp kinks in the phonon dispersion. The two Kohn anomalies originate from a non-analytic behavior of the phonon dispersion, which is impossible to be reproduced by a finite set of interatomic force constants. 
All few-nearest-neighbor force-constant approaches yield a continuous slope at $\Gamma$ and K.

In summary, the divergencies from \emph{ab initio} curves appearing in the high-frequency region are due to the natural limit of accuracy of empirical force-constant models, which consider a finite number of nearest-neighbor atoms and miss the long-range character of the dynamical matrix.

\subsection{Results for carbon nanotubes}
In this section we present phonon dispersion relations of achiral carbon nanotubes that rely on the earlier determined force constants of graphene.
We concentrate in particular on the commonly studied (10,10) and (10,0)~CNT. 
The calculated phonon spectrum of a (10,10) CNT is illustrated in Fig.~\ref{fig:10,10CNT} for both the original parametrization of Saito \emph{et al.}~\cite{Saito1998} and our new parametrization.
%
\begin{figure}[t]
\centerline{\includegraphics[width=7.7cm]{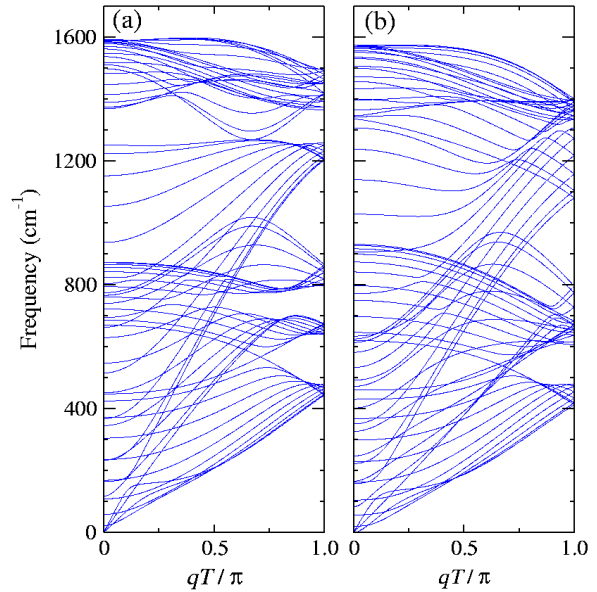}}
\caption{
Phonon dispersion for a (10,10)~CNT calculated with the fourth-nearest-neighbor model with: (a) The parametrization of Saito \emph{et al.}; (b) Our new parametrization. The corresponding parameters are listed in Table~\ref{tab:dr1}. Both parametrizations have been subsequently corrected in order to obtain $\omega=0$ at $q=0$ for the acoustic TW mode.
\label{fig:10,10CNT}
}
\end{figure}
%
In Ref.~[\onlinecite{Saito1998}] was proposed a scaling for the force-constant parameters in order to treat the curvature effect when rolling up the graphene sheet to form a nanotube. This scaling prevents from obtaining a wrong shift of the rotational acoustic mode (TW) at $q=0$ from about $\approx 4$~cm$^{-1}$. In the present work we do not apply the same rescaling, but vary empirically only the out-of-plane tangential force constants $\tilde\phi_{to}^{(n)}=\phi_{to}^{(n)}\,(1+\varepsilon^{(n)})$ of graphene.
These are responsible for vibrations perpendicular to the atom-bonding plane and thus are the most subjected to changes when rolling a plane sheet into a cylinder. This effect increases with decreasing tube diameter.
For a (10,10) nanotube and Saito's parametrization we obtain a frequency of the twisting mode of $\omega\simeq 10^{-3}$ cm$^{-1}$ at $q=0$ by varying only $\phi_{to}^{(4)}$.~\cite{Adaptation}
At this point it is important to observe that even very small variations in the force constants can have considerable effects on the low-frequency modes. In particular the frequency of the quadratically-dispersing modes are strongly affected by modifications of the parameters and can even become imaginary.
For this reason, with our parameter set it was not sufficient to correct $\phi_{to}$ only for the fourth nearest neighbors, but also for third and second neighbors.~\cite{Adaptation} For the latter the correction is smaller, because these are less affected by the effect of curvature. 
It results in a frequency of the TW mode $\omega\simeq 10^{-1}$ cm$^{-1}$ at $q=0$.

\begin{figure}[t]
\centerline{\includegraphics[width=6.8cm]{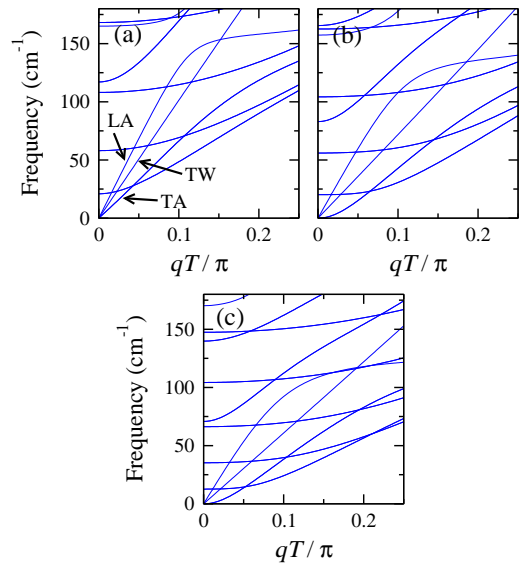}}
\caption{
Low-frequency region of the phonon dispersion relation of a (10,10)~CNT, shown near the $\Gamma$ point, calculated within: (a) 4NNFC model with the original parametrization, (b) 4NNFC model with our new parametrization, (c) Force-constant model by Mahan and Jeon~\cite{Mahan2004} based on three free parameters. While in (a) all four acoustic branches have linear dispersion for small wave vectors, in (b) and (c) two increase linearly with \emph{q} (LA, TW), and two are degenerate (TA) and increase quadratically (TA).
\label{CNT:zoom}
}
\end{figure}

The condition of infinitesimal rotational- and translational invariance imposed on the force-constant tensors gives rise to four zero-frequency modes at $q$~=~0.
Near the $\Gamma$ point the highest-energy acoustic mode is the longitudinal (LA) mode, followed, in order, by a twisting or torsional mode (TW) and a doubly degenerate transverse or flexure mode (TA).
Figure~\ref{CNT:zoom} shows in detail the low-energy region of the phonon spectrum of a (10,10)~CNT for three different cases: (a) and (b) are calculated with the 4NNFC model with the parametrization for graphene of Saito \emph{et al.}~\cite{Saito1998} and with our parametrization, respectively, both corrected for nanotubes. Panel (c) is calculated with a three-parameter spring-and-mass model for carbon nanotubes presented by Mahan and Jeon~\cite{Mahan2004} and is shown for direct comparison. 
The 4NNFC model with the original parametrization of Saito \emph{et al.} shows a linear dispersion at small wave vectors for all four acoustic modes. With our new parametrization of the 4NNFC model, adapted to nanotubes, we found that, while the high-energy optical phonons do not vary significantly, remarkable changes occur for the acoustic modes. The two degenerate TA modes now show quadratic dispersion near the zone center, which was not given by Saito's parametrization.
The model of Mahan and Jeon also obtains the quadratic behavior of the flexure modes, due to a detailed analysis and implementation of symmetry rules, which are required by the correct force constants.
%
%

Also in case of nanotubes with other diameters and chiralities we obtain the correct quadratic dispersion when applying our new parametrization. Figure~\ref{fig:10,0CNT} shows the phonon dispersion of a (10,0)~CNT calculated with the 4NNFC model: Saito's parametrization provides linear dispersions for all four acoustic modes (panel (a)-(c)), while with our parametrization we obtain the quadratic TA mode (panel (b)-(d)), which is given also by the model of Mahan and Jeon (panel (e)).
%
\begin{figure}[t]
\centerline{\includegraphics[width=6.9cm]{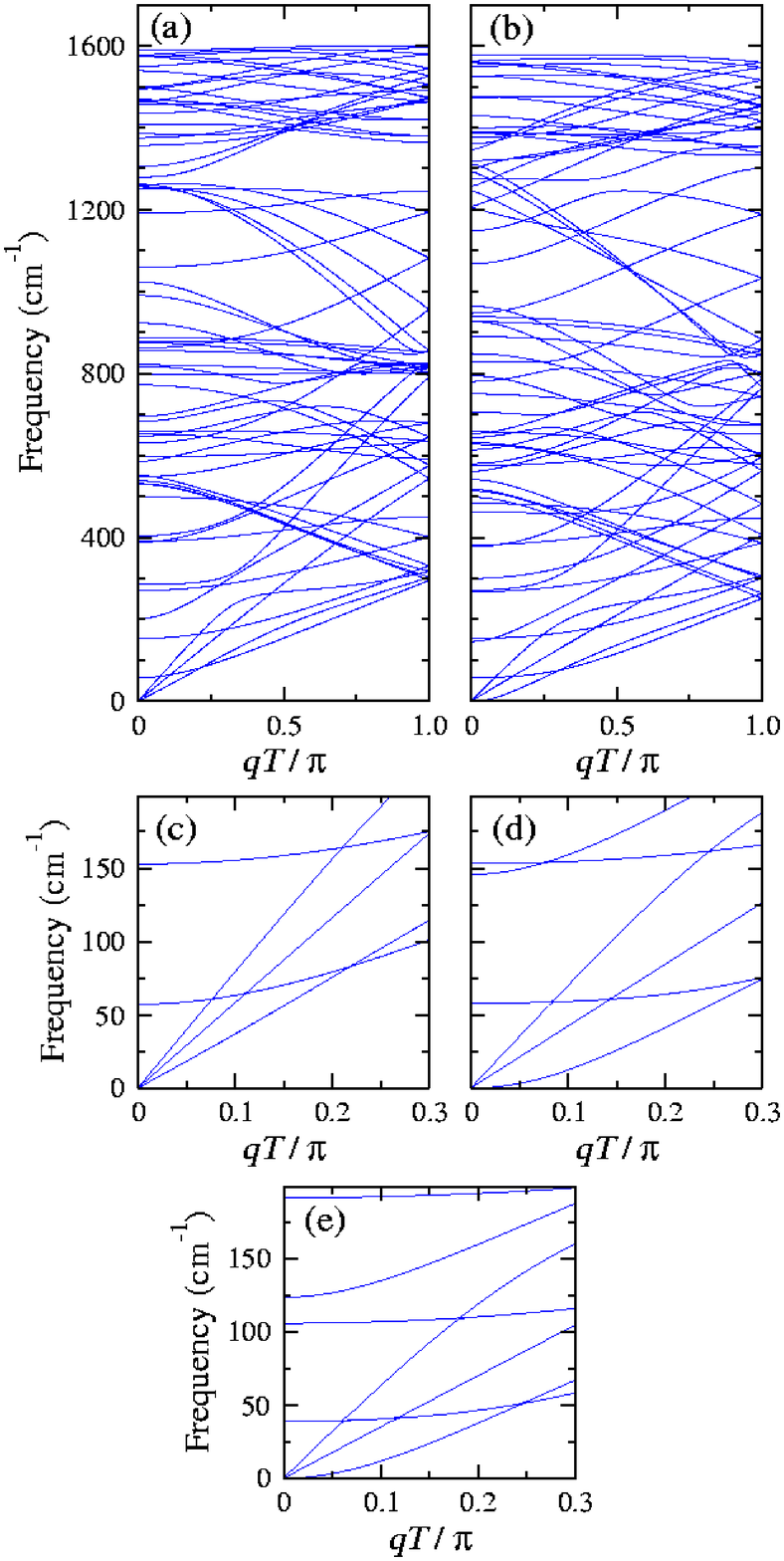}}
\caption{
Phonon dispersion for a (10,0)~CNT: (a) and (c) 4NNFC model with the parametrization of Saito \emph{et al.}; (b) and (d) 4NNFC model with our new parametrization. The two parametrizations show respectively linear and quadratic dispersion for the TA mode. For both the $\phi_{to}$ constants have been corrected in order to obtain zero frequency for the acoustic TW mode at the Brillouin-zone center. (e) Low-frequency region of the spectrum, calculated with the model of Mahan and Jeon, which gets the correct quadratic dispersion of the TA mode. 
\label{fig:10,0CNT}
}
\end{figure}
%
In particular, after adapting the force constants $\phi_{to}^{(n)}$,~\cite{Adaptation} as in the case of the (10,10) CNT, we obtain $\omega\simeq 10^{-1}$ cm$^{-1}$ at $q=0$ for the TW acoustic mode, for both parametrizations.

Furthermore, we concentrate on the important Raman-active radial breathing mode (RBM). This mode arises from a radial expansion and contraction of the entire tube. It is unique to single-walled CNTs and plays an important role in experiments.~\cite{Rao1997,LeRoy2004}
One of the most important applications of the RBM is the determination of nanotubes diameters on the basis of Raman data, through the expected dependence of the RBM frequency on diameter
%
\begin{equation}
\omega_{\textrm{RBM}}=\frac{C_1}{d^{\kappa}_t}+C_2(d_t)
\label{Eq:RBM}
\end{equation}
%
where $C_1$ is a constant, $C_2$ possibly depends on the diameter $d_t$ and $\kappa$ is an exponent. This functional dependence was first introduced by Jishi \emph{et al.}~\cite{Jishi1993} with $C_2$=0 and $\kappa$=1.
Several articles and a range of values of $C_1$ have been published, differing from each other by a few per cent. A review of the experimental and theoretical values can be found in Ref.~[\onlinecite{ReichTM04}]. For isolated tubes the values ranges from $C_1=218$ to $248$ cm$^{-1}$\,nm.

\begin{figure}[t]
\centerline{\includegraphics[width=6.6cm]{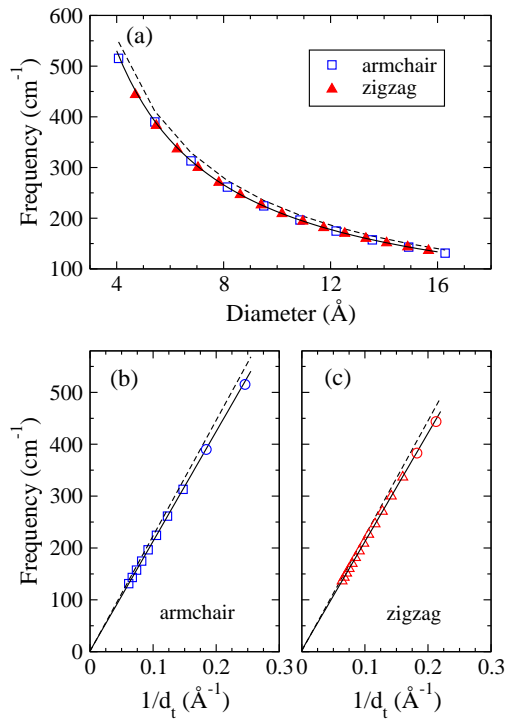}}
\caption{
(a) Frequency of the radial-breathing mode of various armchair ($n$~=~3~$-$~12) and zigzag ($n=6-20$) tubes as a function of the nanotube diameter, calculated with the 4NNFC model with our new parametrization. For comparison, the dashed lines show the results for the original parametrization of Saito. (b) and (c) Frequency of the RBM as a function of the inverse tube diameter for armchair and zigzag tubes, respectively. The solid lines are a linear fit to the data excluding the small-diameter tubes (3,3), (4,4), (6,0), and (7,0), which are marked by circles. These show a deviation from the predicted behavior, with a decrease in the RBM frequency. Reference~[\onlinecite{Rubio1999}] explains it as a consequence of the hybridization changes and the decrease of the $\pi$ interaction induced by the curvature.
\label{fig:RBM}
}
\end{figure}

We verify the relation of Eq.~(\ref{Eq:RBM}) and analyze the chirality dependence of the RBM frequency.
For this purpose we calculate the RBM frequencies of a number of armchair and zigzag nanotubes with the 4NNFC model with our parametrization adapted to CNTs. The obtained frequencies are almost perfectly inverse proportional to the radius of the tube, as shown in Fig.~\ref{fig:RBM}a, and independent on chirality. 
%
%
The RBM frequency decreases with increasing tube diameter and becomes zero in the limit of infinite diameter, which corresponds to the out-of-plane tangential acoustic mode of graphene at $q=0$.
By fitting the frequencies of the RBM to tube diameters by the relation of Eq.~(\ref{Eq:RBM}), we get negligible values for $C_2$ (order of 10$^{-1}$\,cm$^{-1}$\,nm), and $C_1=212$\,cm$^{-1}$\,nm for armchair tubes (Fig.~\ref{fig:RBM}b), which is in satisfactory agreement with the experimental value of 224\,cm$^{-1}$\,nm.~\cite{Rao2001} For zigzag nanotubes (Fig.~\ref{fig:RBM}c) we obtain $C_1=209$\,cm$^{-1}$\,nm. The values of $C_1$ are in agreement also with previous calculations.~\cite{Popov1999,Rubio1999,Kuzmany1998} 
The frequencies obtained by the original parametrization of Saito \emph{et al.} are displayed in Fig.~\ref{fig:RBM} by the dashed lines. We obtain $C_1=223$\,cm$^{-1}$\,nm for armchair tubes and $C_1=222$\,cm$^{-1}\,$nm for zigzag nanotubes. They are in better agreement with experimental values than our parametrization, since the RBM frequencies are on average about 8\,cm$^{-1}$ higher.

A possible chirality dependence is below the resolution of the data. Indeed the proportionality constant $C_1$ differs only by about 1\% between armchair and zigzag nanotubes.
This can be explained by the fact that the RBM corresponds to a stretching of the graphene sheet in the [110] (armchair tubes) or [100] (zigzag tubes) direction. Because the system is isotropic in the hexagonal plane, the elastic constant that the describes the stretching of a graphene sheet is independent on the direction.~\cite{Kuzmany1998} 

Beneath the RBM, also the other low-frequency modes depend strongly on the tube diameter.~\cite{Saito1998,Liu2004} Instead, the higher-frequency modes do not have such a strong diameter dependence since their frequencies are more sensitively determined by the local displacements of the atoms.

\section{Low-temperature thermal properties}
\subsection{Methods}
\subsubsection{Specific heat}
In order to characterize the specific heat at constant volume of low-diameter single wall CNTs, we start from the definition
%
\begin{equation}
c_{\textrm{V}}  =  \frac{1}{V}\left (\frac{\partial E}{\partial T} \right ) _V
\end{equation}
%
where the internal energy $E= k_{\textrm{B}}T^2 (\partial \textrm{ln}Z/\partial{T})$ 
is defined through the vibrational partition function $Z$.
According to statistical thermodynamics, the partition function of a system of independent harmonic oscillators 
can be directly expressed in terms of the phonon frequencies by
%
\begin{equation}
Z=\prod_{\mathbf q, s}\frac{\textrm{e}^{-\hbar\omega_s(\mathbf q)/2k_{\textrm{B}}T}}{1-\textrm{e}^{-\hbar\omega_s(\mathbf q)/k_{\textrm{B}}T}}.
\end{equation}
%
Here, $\mathbf{q}$ is the phonon wave vector, 
$\omega_s(\mathbf{q})$ are the phonon frequencies with mode index $s=1,\ldots,3r$ ($r$ is the number of atoms per unit cell, see Sec.~\ref{Sec:latticedyn}), $T$ is the temperature, $k_{\textrm{B}}$ the Boltzmann constant and $\hbar$ the Planck constant.  
The specific heat can thus be written as
%
\begin{equation}
c_{\textrm{V}} =  \frac{k_{\textrm{B}}}{V}\sum_{\mathbf q,s}\left (\frac{\hbar\omega_s(\mathbf q)}{2k_{\textrm{B}}T}\right )^2\frac{1}{\sinh^2 (\hbar\omega_s(\mathbf q)/2k_{\textrm{B}}T) }.
\end{equation}
%
We do not distinguish between specific heat at constant volume, $c_{\textrm{V}}$, or constant pressure, $c_{\textrm{P}}$, since the model approaches do not include thermal expansion of the lattice. Anyway, the difference between $c_{\textrm{V}}$ and $c_{\textrm{P}}$ is only in the range of a few percent.~\cite{Mounet2005}
For densely spaced values of the wave vector $\mathbf{q}$ it is possible to replace the sum by an integral
%
\begin{eqnarray}
\sum_{\mathbf q,s}\rightarrow \sum_s\int \textrm{d}\mathbf q & \equiv & \sum_s \frac{V}{(2\pi)^3}\int \textrm{d}^3 q
\\
& = & 3rN \int_0^\infty g(\omega)\textrm{d}\omega
 \end{eqnarray}
%
where $g(\omega)$ is the density of states (DOS) and $N$ is the number of unit cells. The expression for the specific heat results
%
\begin{equation}
c_{\textrm{V}}=3rk_{\textrm{B}}\int_0^\infty \textrm{d}\omega \left ( \frac{\hbar\omega}{2k_{\textrm{B}}T}\right )^2 \frac{g(\omega)}{\sinh^2 (\hbar\omega/2k_{\textrm{B}}T)}.
\label{Spec:heat}
\end{equation}
%
Therefore the specific heat depends in a detailed way on the frequency spectrum $g(\omega)$ of the normal modes.

\subsubsection{Landauer phonon transport}
Phonon heat transport in mesoscopic systems can be investigated using methods analogous to the Landauer description of electrical conductance.~\cite{Rego1998,Segal2003} We consider a model of an ideal one-dimensional heat conductor, built by two long perfect leads that join a central segment in which the phonon scattering occurs. Only elastic scattering is taken into account, while phonon-phonon interaction is neglected. The free ends of the two leads are connected to reservoirs of temperature $T_{\textrm{hot}}$ and $T_{\textrm{cold}}$, respectively. No scattering occurs at the reservoir-lead connections. The energy flux of the right/left moving phonons is given by~\cite{Rego1998,Schwab2000,Angelescu1998}
%
\begin{eqnarray}
J^{+/-} & = & \frac{1}{2\pi}\sum_s \int_0^\infty \textrm{d}q\,\hbar\omega_s(q)\, \eta_{\substack{\textrm{hot}/\\\textrm{cold}}}(\omega_s(q)) \, v_s(q) \,\mathcal T_s(q)\nonumber \\
& = & \frac{1}{2\pi}\sum_s \int_{\omega_s^{\textrm{min}}}^{\omega_s^{\textrm{max}}} \textrm{d}\omega\,\hbar\omega\, \eta_{\substack{\textrm{hot}/\\\textrm{cold}}}(\omega)\, \mathcal T_s(\omega)
\end{eqnarray}
%
where $\omega_s(q)$ is the dispersion relation of the discrete mode $s$, $v_s(q)$ is the group velocity and $\mathcal T_s(q)$ are transmission coefficients characterizing the coupling of wave\-guide modes to the reservoirs. The total heat current is therefore $J_{\textrm{ph}}=J^+-J^-$. 
Assuming perfectly adiabatic contact between the thermal reservoirs and the ballistic quantum wire, the transmission function for a monotonically dispersing mode $s$ is the step function
%
\begin{equation}
\mathcal T_s(\omega)= \begin{cases}
	~1 & ~~~\textrm{for} ~~\omega_s^{\textrm{min}}\leq \omega \leq \omega_s^{\textrm{max}}, \\
	~0 & ~~~\textrm{otherwise}.
     \end{cases}
\label{T:step:function}
\end{equation}
%
Instead, for non-monotonic dispersions, given a frequency $\overline{\omega}$,
the transmission $\mathcal T_s(\overline{\omega})$ is defined as the number of crossings of the line $\omega=\overline{\omega}$ with the phonon dispersion of the mode~$s$.
With the total transmission function given by $\mathcal T=\sum_s \mathcal T_s(\omega)$, the Landauer energy flux results~\cite{Rego1998}
%
\begin{equation}
J_{\textrm{ph}}=\int_0^{\infty} \frac{\textrm{d}\omega}{2\pi}\,\hbar\omega \, [\eta_{\textrm{hot}}-\eta_{\textrm{cold}}] \, \mathcal T(\omega).
\label{Landauer:current:2}
\end{equation}
%
Eventually, the thermal conductance is defined as
%
\begin{equation}
\kappa_{\textrm{ph}}=\frac{J_{\textrm{ph}}}{\Delta T}
\end{equation}
%
with $\Delta T=T_{\textrm{hot}}-T_{\textrm{cold}}$.
In the limit of linear response, $\Delta T\ll T\equiv(T_{\textrm{hot}}+T_{\textrm{cold}})/2$, we obtain using Eq.~(\ref{Landauer:current:2}) and the substitution $x=\hbar\omega/k_{\textrm{B}}T$
%
\begin{equation}
\kappa_{\textrm{ph}}=\frac{k_{\textrm{B}}^2T}{h}\int_0^\infty\textrm{d}x \, \frac{x^2\textrm{e}^x}{(\textrm{e}^x-1)^2} \, \mathcal T\left ( x\frac{k_{\textrm{B}}T}{\hbar}\right ).
\label{Therm:cond}
\end{equation}
%
This equation plays the role of a `universal' phonon conductance.
An important statement is that the result is independent of all details of the dispersion curve except the transmission function. This arises because the density of states in the frequency integral is canceled by the group velocity.

\subsection{Specific heat results}
The specific heat of carbon nanotubes is mainly determined by phonons, while electronic contributions to it can be neglected even at a few Kelvin.~\cite{Benedict1996}
According to Eq.~(\ref{Spec:heat}) it depends sensitively on the characteristics of the phonon spectrum and on its vibrational density of states (DOS). 

\begin{figure}[t]
\centerline{\includegraphics[width=6.5cm]{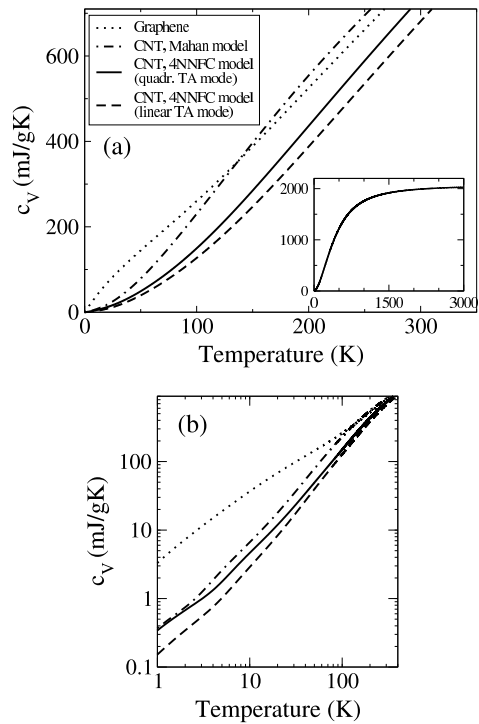}}
\caption{(a) Specific heat as a function of temperature for graphene (dotted line), calculated with the 4NNFC model and our new parameters, and for a (10,10) CNT calculated with the model of Mahan-Jeon (dot-dashed line), the 4NNFC model with our parameters (solid line), and the 4NNFC model with the original constants of Saito \emph{et al.} (dashed line). The inset shows a wider temperature interval: $c_{\textrm{V}}$ approaches the value 2078 mJ/gK for high temperatures. (b) Specific heat on a logarithmic scale for the low frequency region. The assignment of the line styles is the same as in (a).
\label{C_v1}
}
\end{figure}

The specific heat calculated from the theoretical DOS spectra is shown in Fig.~\ref{C_v1} as a function of temperature.
At high temperatures, the specific heat of all the different approaches and all chiralities converges to the classical limit of $3k_{\textrm{B}}/M=2078$ mJ/gK with $M$ being the atomic mass of carbon (see inset in Fig.~\ref{C_v1}a).
In the low-temperature regime that we are interested in (below 20~K, see Fig.~\ref{C_v1}b), the specific heat of graphene is dominated by the quadratic out-of-plane bending mode and is expected to have a linear $T$ dependence at very low temperatures. The nanotube curve is lower than the graphene one because the tube has no low-energy counterpart to the layer-bending modes.~\cite{Hone2000} 
In this temperature range, nanotube modes with $\hbar\omega_s(\mathbf q)\gg k_{\textrm{B}}T$ will negligibly contribute to Eq.~(\ref{Spec:heat}), since the integrand will vanish exponentially. Optical phonons are frozen out and only the long-wavelength acoustic modes are populated, because for these holds $\omega_s(q)\rightarrow 0$ as $q\rightarrow 0$. 
Therefore, the acoustic modes alone determine the low-temperature behavior of the specific heat. Only at a temperature $T_{\textrm{opt}}\approx\hbar\omega_{\textrm{opt}}/6k_{\textrm{B}}$ does the lowest-lying optical subband with frequency  $\omega_{\textrm{opt}}$ at $q=0$ begin to contribute to the specific heat.~\cite{Mizel1999}
%
%
The two $c_{\textrm{V}}$ curves calculated by the model of Mahan (dot-dashed line in Fig.~\ref{C_v1}b) and by the 4NNFC model with our parametrization (solid line) coincide in the temperature range of a few Kelvin. This was expected because both models predict quadratically dispersing flexure modes and thus analogous low-temperature behavior.
The slope of the curves increases smoothly when the first optical subband begins to contribute to $c_{\textrm{V}}$. In the model of Mahan this takes place approximately at $T_{\textrm{opt}}\approx$3~K, while for our parametrization it appears about 5~K, due to the different frequencies of $\omega_{\textrm{opt}}$, that are 12.6 cm$^{-1}$ and 20.2 cm$^{-1}$, respectively.
The 4NNFC model with Saito's parametrization instead yields linearly dispersing flexure modes and the $(\textrm{log}\,c_{\textrm{V}})$ vs.\ $(\textrm{log}\,T)$ curve (green line in Fig.~\ref{C_v1}b) shows a higher slope than the two curves described just now. Due to the lowest lying optical mode with $\omega_{\textrm{opt}}$ = 21.0 cm$^{-1}$, the slope increases at $\sim$5~K, as expected.

The low-temperature behavior of the $c_{\textrm{V}}$ vs. $T$ curves needs to be analyzed in more detail.
Indeed, the behavior of $c_{\textrm{V}}$ contains informations regarding the dimensionality of the system through a fixed correlation between $c_{\textrm{V}}$, the DOS and the exact dispersion law.~\cite{Popov2002} According to Table~\ref{DOS:cv}, through evaluation of the exponent $\alpha$ in the power law $c_{\textrm{V}} \varpropto T^{\alpha}$ it is possible to get informations about the dimensionality of the system. Since nanotubes are quasi-one dimensional (1D) systems consisting of rolled-up 2D sheets, they should display both 1D quantum size effects and 2D features.
%
\begin{table}[t]
\begin{tabular}{c c c c}
\hline
\hline
Dimension & Phonon  & Phonon & Specific \\
	  & dispersion	& DOS & heat \\
\hline
1D & $\omega\varpropto q^2$ & $g(\omega)\varpropto1/\sqrt{\omega}$ & $c_{\textrm{V}} \varpropto \sqrt{T}$ \\
   & $\omega\varpropto q$ & $g(\omega)=\textrm{const}$ & $c_{\textrm{V}} \varpropto T$ \\	
\hline
2D & $\omega\varpropto q^2$ & $g(\omega)=\textrm{const}$ & $c_{\textrm{V}} \varpropto T$ \\
   & $\omega\varpropto q$ & $g(\omega)\varpropto\omega$ & $c_{\textrm{V}} \varpropto T^2$ \\
\hline
3D & $\omega\varpropto q$ & $g(\omega)\varpropto\omega^2$ &$c_{\textrm{V}} \varpropto T^3$ \\
\hline
\hline
\end{tabular}
\caption{
Low-temperature behavior of the specific heat. The dimensionality of the system is correlated to the density of states and, therefore, to the specific heat. At low temperature only acoustic modes are excited. These can have either linear or quadratic dispersion. 
\label{DOS:cv}}
\end{table}
%

\begin{figure}[t]
\centerline{\includegraphics[width=7.5cm]{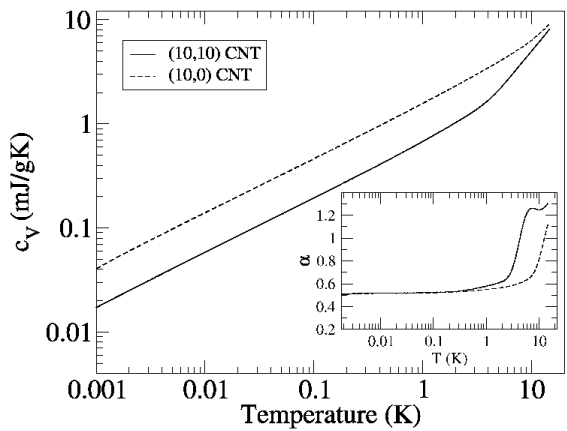}}
\caption{ 
Low-temperature specific heat for a (10,10) and a (10,0) CNT calculated with the 4NNFC and our new parametrization. The inset shows the value of their slope $\alpha =\textrm{d}(\textrm{log}\,c_{\textrm{V}})/\textrm{d}(\textrm{log}\,T)$.
\label{expon}
}
\end{figure}

Figure~\ref{expon} shows the low-temperature specific heat for the (10,10) and the (10,0) with the 4NNFC model and our parametrization, and the inset shows the slope $\alpha$ of  $(\textrm{log}\,c_{\textrm{V}})$ vs.\ $(\textrm{log}\,T)$ curves.
In the mK temperature range $\alpha$ clearly tends to the value $1/2$. This is almost entirely due to the doubly degenerate flexure mode with quadratic dispersion law (see Table~\ref{DOS:cv}), which is dominating in this temperature range.
With increasing temperature (0.8 to 5~K for the (10,10) CNT), also the contribution of the two linearly dispersing modes becomes stronger and $\alpha$ holds values between 1/2 and 1 due to the superposition of these four modes. 
Above $T_{\textrm{opt}}$ the slope changes considerably due to the optical phonons and the tube is essentially 2D. 
%
%
This behavior is in accordance with theoretical predictions~\cite{Hone2000} and is a direct confirmation of quantized 1D phonon subbands in carbon nanotubes.
Our results for low-temperature $c_{\textrm{V}}$ agree very well with previous calculations of Popov,~\cite{Popov2002} and additionally we extended the low-temperature limit by two orders of magnitude. We find good agreement also with experimental results of Lasjaunias~\emph{et al.},~\cite{Lasjaunias2002} who measured the specific heat down to 0.1~K and fitted their measured curves with a power law of $0.043T^{0.62}+0.035T^3$. However, experimental measurements are usually performed on bundles of nanotubes, whose properties can differ greatly from those of isolated tubes. The adding of tubes to a bundle suppresses in particular the bending flexure modes, with a consequent increase of the exponent $\alpha$ in favor of a linear $T$ dependence.

\begin{figure}[t]
\centering
\includegraphics[width=7.9cm]{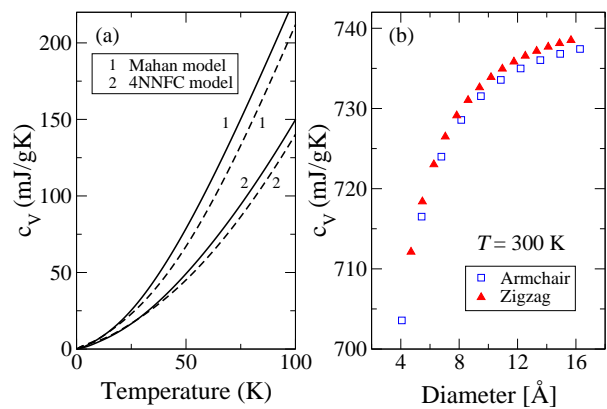}
\caption{Chirality dependence of the specific heat: (a) Temperature dependence of $c_{\textrm{V}}$ for a (10,10) CNT (straight lines) and a (10,0) CNT (dashed lines). The index 1 refers to the model of Mahan-Jeon and the index 2 to the 4NNFC model with our parameters; (b) The effect of tube diameter on zigzag and armchair CNT specific heat. At a given temperature the specific heat increases with the increase of tube diameter. The upper limit is given by graphene, with $c_{\textrm{V}}=794$ mJ/gK at 300~K. \label{C_v2}}
\end{figure}

We achieved similar results for other armchair and zigzag nanotubes. However, since the first optical subband edge varies from tube to tube and depends on the model, the turning points in $c_{\textrm{V}}$ vs. $T$ curves are different, causing a crossover of $c_{\textrm{V}}$ curves. The general uptrend and the high-temperature limit are the same. Figure~\ref{C_v2}a shows the specific heat curves for a (10,10) and a (10,0) CNT.
%
%
The tube diameter influences the specific heat of carbon nanotubes, especially in the range of 25-350~K.
In order to determine the effect of tube diameter on the specific heat, additional results for $T=300$~K are displayed in Fig.~\ref{C_v2}b, using the 4NNFC model with our parametrization. 
At a fixed temperature the specific heat increases with increasing tube diameter. This was as expected, since for very large diameters the curve should approach the $c_{\textrm{V}}$ value of graphene, which is 794~mJ/gK at 300~K. However, the effect decreases at large tube diameter. 
The chirality shows only a small effect in the tubes specific heat, with $c_{\textrm{V}}$ of the zigzag tubes lying over that of the armchair tubes. This small effect is negligible and could also be caused by inaccuracies of the model description. The results are in good agreement with those of Ref.~[\onlinecite{Dobardzic2003,Li2005}].

\subsection{Thermal conductance}
\begin{figure}[t]
\centering
\includegraphics[width=6.9cm]{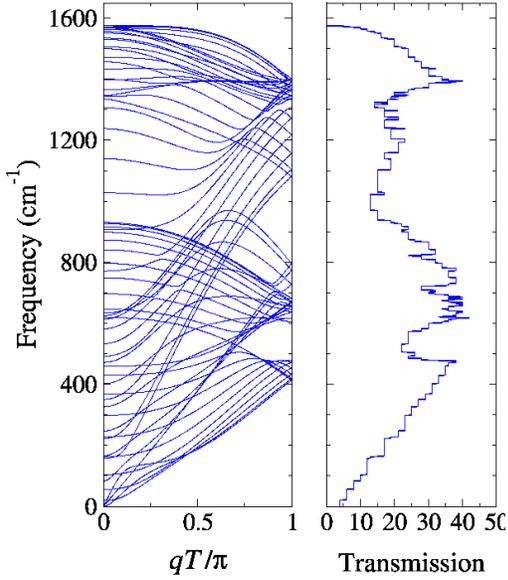}
\caption{
Phonon dispersion of a (10,10) CNT and the transmission function $\mathcal T$~=~$\sum_s \mathcal T_s(\omega)$. The latter is a sum of $s=120$ step functions. \label{Transm}}
\end{figure}

In the following, we demonstrate that at low temperatures a carbon nanotube behaves as a ballistic, one-dimensional wire and that the phonon thermal conductance is quantized.
The thermal conductance can be calculated by evaluating Eq.~(\ref{Therm:cond}).
In this expression, the integrand is given by the product of two functions: the transmission function and a weight function $x^2\textrm{e}^x/(\textrm{e}^x-1)^2$.
The former is related to the phonon spectral properties of the nanotube and the latter takes into account the effects due to temperature. Figure~\ref{Transm} shows the transmission function $\mathcal T=\sum_s \mathcal T_s(\omega)$ for a (10,10)~CNT. Some branches count doubly because of their degeneracy.
%
%
For high temperatures, these two functions are non-zero within the same range, which means that all the transmission modes contribute to the thermal conductance.
Whereas, in the limit of low temperature, the broadness of the weight function is extended only to the low-energy modes. Therefore, at low temperatures only four acoustic modes give an appreciable contribution to the thermal conductance of a carbon nanotube.
In this temperature regime Eq.~(\ref{Therm:cond}) becomes greatly simplified
%
\begin{equation}
\kappa_{\textrm{ph}}\simeq\frac{k_{\textrm{B}}^2T}{h} \,4 \int_0^\infty\textrm{d}x \, \frac{x^2\textrm{e}^x}{(\textrm{e}^x-1)^2} = 4 \,\frac{\pi^2 k_{\textrm{B}}^2T}{3h}.
\label{ptc}
\end{equation}
%
Here, the factor 4 represents the number of acoustic modes. The upper limit of the integral is of few importance, because the integrand function falls off rapidly, before the successive step in the transmission function takes place. From Eq.~(\ref{ptc}) results that a fundamental relation holds for each mode
%
\begin{equation}
\kappa_0=\frac{\pi^2 k_{\textrm{B}}^2T}{3h}.
\end{equation}
%
This quantum of thermal conductance represents the \emph{maximum possible value of energy transported per phonon mode}.
It does not depend on particle statistics, therefore, is universal for fermions, bosons, and anyons.~\cite{Rego1999} Furthermore, it is independent of any material parameters and of precise details of the dispersion law. This is clear since to construct the transmission function as in Eq.~(\ref{T:step:function}), it does not matter whether the dispersions are linear or quadratic, but the branch upper and lower limits should be accurately computed.

\begin{figure}[t]
\centering
\includegraphics[width=7.5cm]{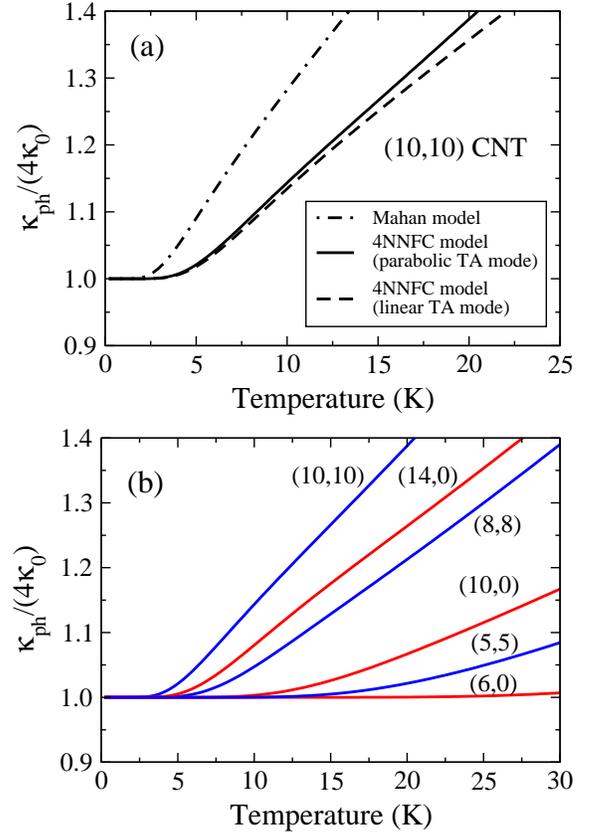}
\caption{(a) Phonon thermal conductance for a (10,10) CNT calculated with the Mahan-Jeon model (dot-dashed line), the 4NNFC model with our parameters (solid line), and the 4NNFC model with the original constants of Saito \emph{et al.} (dashed line). The two latter show, respectively, quadratic and linear dispersion for the TA mode. (b) Thermal conductance for several carbon nanotubes, calculated with our parametrization. \label{Conductance}}
\end{figure}

The phonon thermal conductance $\kappa_{\textrm{ph}}$ of a (10,10)~CNT as a function of temperature is shown in Fig.~\ref{Conductance}a, already normalized by $4\kappa_0$. For temperatures in the range of a few Kelvin, this ratio reaches the constant value of 1, independent of the model approach and of chirality.
This behavior confirms that the thermal conductance of carbon nanotubes is quantized.
Despite the quantization, the curves do not present steps because of the broadening of the Bose-Einstein distribution in comparison with the energy gap between subband edges.
The length of the plateau depends on the lowest optical frequency of the dispersion curve. Indeed, the turning point in the curve of the (10,10) CNT calculated with the model of Mahan (Fig.~\ref{Conductance}a, dot-dashed line) is about 2K, while it is higher for the 4NNFC model with both parametrizations (about 3K).
As predicted, the exact dispersion law of the acoustic TA mode does not affect the qualitative behavior of the $\kappa_{\textrm{ph}}$ vs. $T$ curve. The two curves for the 4NNFC model result respectively from a quadratic and linear dispersion of the TA mode, but show the same low-temperature behavior. The deviations above 10~K are due to small differences in the respective optical frequencies.
%
%

The results of the 4NNFC model are believed to be more accurate than the ones obtained using Mahan and Jeon's method for the phonon dispersion, because the latter does not correctly reproduce the graphene dispersion.
Eventually, the thermal conductance depends only on the tube radius and not on chirality. Results for armchair tubes are very similar to the ones for zigzag tubes, when same diameters are compared.
This arises because the energy $\hbar\omega$ of the lowest-lying optical modes is determined only by the tube radius and decreases approximately according to $\sim 1/R^2$ (see Ref.~[\onlinecite{Watanabe2004}]).
Figure~\ref{Conductance}b shows the thermal conductance for some armchair and zigzag nanotubes.

Our results are in very good agreement with those of Yamamoto \emph{et al.}~\cite{Watanabe2004} and Mingo \emph{et al.}.~\cite{Mingo2005}
Experimental studies were achieved by Schwab \emph{et al.},~\cite{Schwab2000} who observed the quantum thermal conductance in a nanofabricated 1D structure, which behaves essentially like a phonon wave\-guide. 

Finally, we would like to express a word of caution to specify the limits of the Landauer model of heat conduction. It must be stated that it describes an idealized case of ballistic transport through a one-dimensional wave\-guide,
where the phonon transmission
occurs without scattering by defects or scattering at the reservoir-lead connection. 
The wave\-guide and the reservoirs are coupled adiabatically and anharmonicity and phonon-phonon interactions are neglected. 
These conditions are fulfilled only 
at low temperatures, where the phonon mean free path is limited only by the size of the system and anharmonic terms are small compared with the harmonic part of the Hamiltonian.
However, as far as the system size exceeds the mean free path, which strongly depends on temperature, scattering of phonons due to anharmonic terms of the interatomic potential begins to decrease the conductivity and the transport ceases to be ballistic. 
Indeed, anharmonic terms 
give rise to phenomena as finite phonon lifetimes and interaction between phonons. These determine strongly the transport properties at higher temperatures and are responsible for finite thermal conductivity.


\section{Conclusions}
We presented a combined theoretical investigation of both vibrational and thermal properties of graphene and carbon nanotubes within a force-constant model.
First we fitted the phonon dispersion of graphene to that obtained with \emph{ab initio} calculations by Bohnen and Heid~\cite{HeidPrivat} and found reasonable agreement for the overall dispersion and good agreement for the acoustic modes. The frequency values at high symmetry points $\Gamma$, M, and K lie close to those obtained by various first-principles calculations (about 4\%, with the exception of only one mode). 
Then we presented results for the phonon spectra of achiral carbon nanotubes and focused on the low-frequency region. The dispersion of the doubly-degenerate flexure mode shows $\omega\propto q^2$ behavior at long wavelengths, as predicted by several theoretical works. 
Particular attention has been paid to the radial-breathing mode, with a detailed analysis of the frequency- and chirality dependence on the tubes' diameter.
On the basis of the so-obtained phonon spectra, we calculated the specific heat and the thermal conductance of carbon nanotubes. The quadratic dispersion of the flexure modes leads to a $\sqrt T$ dependence of the specific heat at very low temperatures. This is a direct confirmation of the one-dimensional behavior of carbon nanotubes at low temperature.
Concerning heat transport, we showed that nanotubes can conduct heat by ballistic phonon propagation.
At low temperatures the thermal conductance for a single phonon channel approaches a maximum value of $\kappa_0=\pi^2 k_{\textrm{B}}^2T/3h$, which is the universal quantum of thermal conductance. We showed that for nanotubes of different diameter and chirality, the thermal conductance reaches the value $4\kappa_0$ for $T\rightarrow0$, where the factor 4 is due to the four acoustic modes of a nanotube.
All our results are in very good agreement with theoretical and experimental data available in literature.

\begin{acknowledgments}
We are in debt to R. Heid and K.-P. Bohnen for providing us with the \emph{ab initio} data used throughout this work.
We acknowledge fruitful discussions with D. Bercioux and A. Donarini.
This work was partially funded by the Volkswagen Foundation
under grant No. I/78 340, by the European Union grant
CARDEQ under contract No. IST-021285-2 and by the
Deutsche Forschungsgemeinschaft (DFG) within the Collaborative
Research Center SFB 689.
\end{acknowledgments}



\begin{thebibliography}{25}
\bibitem[{\citenamefont{Adaptation}(2006)}]{Adaptation}
  \bibinfo{author}{\bibnamefont{For a (10,10) CNT we set (a) for Saito's parametrization $\varepsilon^{(4)}\sim 5.122\cdot 10^{-3}$ and (b) for our parametrization $\varepsilon^{(2)}\sim 9.0\cdot 10^{-3}$, $\varepsilon^{(3)}\sim 9.5\cdot 10^{-3}$, and $\varepsilon^{(4)}\sim 1.42\cdot 10^{-2}$. For a (10,0) CNT we set (a) for Saito's parametrization $\varepsilon^{(2)}\sim 9.0\cdot10^{-3}$, $\varepsilon^{(3)}\sim 9.5\cdot10^{-3}$, and $\varepsilon^{(4)}\sim 3.89\cdot10^{-2}$, and (b) for our parametrization $\varepsilon^{(2)}\sim 9.5\cdot10^{-3}$, $\varepsilon^{(3)}\sim 1.0\cdot10^{-2}$, and $\varepsilon^{(4)}\sim 4.88\cdot10^{-2}$   }}.
\end{thebibliography}
\end{document}